\documentstyle[preprint,aps]{revtex}
\newtheorem{df}{Definition}
\newtheorem{th}{Theorem}
\begin{document}
\draft
\tightenlines
\title{Static Quantum Computation}
\author{Haiqing Wei}
\address{
Department of Physics, McGill University\\
Montreal, Quebec, Canada H3A 2T8\\
{\rm E-mail: dhw@physics.mcgill.ca}
}
\author{Xin Xue}
\address{
Department of Natural Resource Sciences\\
Macdonald Campus of McGill University\\
Ste-Anne-de-Bellevue, Quebec, Canada H9X 3V9\\
{\rm E-mail: xkhz@musicb.mcgill.ca}
}
\maketitle

\begin{abstract}
Tailoring many-body interactions among a proper quantum system endows it 
with computing ability by means of static quantum computation in the sense
that some of the physical degrees of freedom can be used to store binary
information and the corresponding binary variables satisfy some given logic
relations if and only if the system is in the ground state. Two theorems are
proved showing that the universal static quantum computer can encode the
solutions for any P and NP problem into its ground state using only
polynomial number (in the problem input size) of logic gates. The second
step is to read out the solutions by relaxing the system. The time
complexity is relevant when one tries to read out the solution by relaxing
the system, therefore our model of static quantum computation provides a
new connection between the computational complexity and the dynamics of a
complex system.
\end{abstract}

\newpage

\begin{center}
{\bf Introduction}
\end{center}

Any computational task is carried out by a certain kind of physical system.
All the existing and proposed computers are dynamic systems. The computation
is accomplished via programmed dynamic evolutions of the physical system. In
a state-of-the-art digital electronic computer, electromagnetic pulses are
used to drive the thermodynamic evolution of the electronic system
that achieves the irreversible computation. In a dynamic (this adjective is
often omitted in the literature) quantum computer [1,2,3], a series of
time-dependent Hamiltonians are applied to realize the unitary operations on
the quantum states. The dynamic quantum computation is carried out in a
reversible manner. In a recent work [4], it has been
described how to construct a quantum search machine to encode the
3-SATISFIABILITY problem which is a famous hard search problem in computer
science [5]. The search machine accomplishes its task of encoding the
solution for a given problem into its ground state by no dynamic
evolution but the so-called static quantum computation (see the definition
below). nspired by this achievement, a natural question to ask is
whether similar machines can be constructed that encode other NP problems
efficiently. In this article, it will be shown that a universal static quantum
computer (see the definition below) can encode the solutions for any P and NP
problem into its ground state using only polynomial number of logic gates.

\begin{center}
{\bf Basic Elements of Static Quantum Computation}
\end{center}

\begin{df}
Any quantum system having physical degrees of freedom $x_1,x_2,\cdots x_n$
and $y$ to which input and output variables are assigned respectively, is
said to evaluate the function
\begin{equation}
y=F(x_1,x_2,\cdots x_n)
\end{equation}
by means of {\sc static quantum computation} provided that the system lies
at the lowest energy level when and only when the assigned variables satisfy
Eq.1. Those states whose variables fail to satisfy Eq.1 are associated with
higher energies. Any quantum system that evaluates logic functions via
static quantum computation is regarded as a {\sc static quantum computer}
(SQC).
\end{df}
The idea of static quantum computation has been implicitly used to make
simple logic gates [6], yet its intimate relation with computational
complexity has not been realized.
Our above definition of SQC is general. However, in order to show
a concrete example about how a static quantum computer can be implemented
as well as how it achieves the computational power, this article will
concentrate on a specific type of SQC in which the so-called binary wire
serves as the basic most constructing element.
\begin{df}
A {\sc binary wire} is a linear chain of strongly interacting atomic bodies
which has two degenerate ground states of collective motion (called
{\sc working mode}s) that can be used to store binary information. By virtue
of its spatial extension, a binary wire can also serve as information
transmission line.
\end{df}
By definition, any linear chain of strongly interacting atomic bodies is a
potential candidate for the binary wire. Examples are linear ferromagnetic or
antiferromagnetic chains, lines of Coulomb interacting quantum dots [7],
linear arrays of interacting superconductors, {\it ect}. Although the binary
wire proposed in Ref.[7] is by no means the only possible implementation, it
is conceptually instructive and almost realizable by today's technology.
\begin{df}
A {\sc static quantum logic gate} (SQLG) evaluating the logic function
$F$ is a device consisting of input and output binary wires which are brought
close to interact and the interaction among them are so tailored that the
total ground state interaction energy is minimized if and only if the output
variable $y$ and the input variables $(x_1,x_2,\cdots,x_n)$ satisfy the
logic relation as given by Eq.1.
\end{df}
Excellent examples of SQLG can be found in Ref.[6]. 
\begin{df}
A {\sc static quantum network} (SQN) is a network of SQLGs connected by
binary wires serving as either binary registers or information
transmission lines.
\end{df}
\begin{df}
An {\sc energy degeneracy conserving} (EDC) {\sc static quantum logic gate}
is an SQLG that for all possible inputs, the ground states of this single gate
always have the same amount of interaction energy. Hence an EDC gate has the
property of energy degeneracy conservation in the sense that when it is
connected to an SQN, it shifts all the possible quantum states of the network
by the same amount, leaving all degenerate states degenerate.
\end{df}
In Ref.[6] there are examples for both EDC and non-EDC SQLGs. When a logic
inverter is connected to any SQN, its ground states always contribute the
same amount of interaction energy in spite of the input. The inverter is EDC
in the sense that it conserves the energy degeneracy among all the states of
the whole network. The dedicated AND gate is a contrast. For two different
inputs $A=0$, $B=0$ and $A=1$, $B=0$ the outputs are the same $C=0$. In each
case the interaction energy inside the AND gate is minimized so that the gate
itself is in the ground state. However the two ground states obviously have
different energies. Hence the dedicated AND gate is non-EDC in the sense
that when connected to an SQN it won't conserve the energy degeneracy.
Fortunately, all non-EDC SQLG can be made EDC by an
{\sc input symmetrization} technique. Since any logic gate can be decomposed
into two-level AND gates and inverters [8], it is sufficient to show that the
two-level AND gate can be made EDC. There are only four possible inputs for
an AND gate with two input binary wires. A single AND gate may contribute
at most four different values of interaction energy. One can construct a
{\sc symmetrized and} (SAND) {\sc gate} as shown in Fig.1 in which four AND
gates exhaust all the four possible situations. For any inputs $A$, $B$ the
ground state energy of the SAND gate is always the sum of the four possible
values of interaction energy inside a single AND gate. Notice that only the
output of one AND gate $C_0$ serves as the output of the SAND gate, other
three outputs $C_1$, $C_2$ and $C_3$ are dangling. The SAND gate is
obviously EDC.
\begin{df}
An {\sc energy degeneracy conserving static quantum network}
(EDCSQN) is a logic network of EDC SQLGs.
\end{df}
To construct the static quantum computer using EDCSQN has the great advantage
which will be seen later.

An SQN evaluates logic functions in the static manner that as soon as the
whole network has been constructed and the inputs have been set, the
evaluation is over. No dynamic evolution is needed. The right answers together
with the corresponding inputs are encoded into the ground state. For this
reason when the idea of static computation was used to achieve simple logic
operations, it has been called ``computing with the ground state'' [6]. One
significant observation is that for each binary wire there are two working
modes representing binary values $1$ and $0$, when $n$ input binary wires
are connected to an EDCSQN, essentially all the integers from $0$ to $2^n-1$
in binary form will be sent to the logic network. The EDCSQN evaluates
a given function for all the $2^n$ inputs in parallel and all the results
are stored in the output binary wires. This gives birth to the notion of
{\sc static quantum parallelism}. By virtue of this parallelism, an EDCSQN
accomplishes nondeterministic computation [5] in the static manner.

\begin{center}
{\bf Static Computational Complexity}
\end{center}

\begin{df}
The {\sc static computational complexity} of an SQC or SQN is the number of
basic static logic elements that it consists of. The static computational  
complexity of a problem with input size $n$ is the least number of basic
static logic elements (as the function of $n$, denoted by $SCC$($n$)) needed
to construct an SQC which encodes the solutions of the problem into its
ground state. The basic static logic elements are the binary wire register,
the logic inverter, the AND and the OR gate.
\end{df}
As in the theory of computational complexity based on the Turing machine,
problems can be classified according to the static computational complexity.
\begin{df}
A problem is in the class SP if its static computational complexity function
$SCC(n)$ is bounded by a polynomial ${\cal P}$, {\it i.e.}
\begin{equation}
SCC(n)\le {\cal P}(n)
\end{equation}
\end{df}
Notice that unlike the definition of the well-known class P on Turing
machine, problems in the class SP are not restricted to be deterministic.

According to the classical definition, a problem in class P with input size
$n$ can be solved in $p(n)$ steps of computation on a deterministic one-tape
Turing machine (DTM), which consists of a finite state control, a read-write
head, and a tape made up of a two-way infinite sequence of tape squares [5]. A
program runs on the DTM in a step-by-step manner. In each step the read-write
head may read the symbol in the tape square under it, then the finite state
control changes its own state, gives a new symbol to the read-write head which
in turn erases the old symbol in the tape square and writes the new one into
it. Also the finite state control gives the read-write head an instruction to
move one square left or right. Using SQLGs, a static quantum DTM (SQDTM) in
analogue to the classical DTM can be constructed. The construction starts with
binary wire registers to simulate the tape of the DTM. Upon the problem in
class P being solved on the DTM, the number of tape squares which
have ever been scanned by the read-write head is no more than the number of
steps of computation, $p(n)$, because in each step the head scans only one
square. For each step of the DTM computation, a register consisting of
$p(n)$ bare binary wires is enough to embody the ever-scanned tape squares in
the sense that each square has a corresponding binary wire to simulate it.
The binary wires in a register are labeled by an integer $j$,
$j=1,2,\cdots ,p(n)$.
Totally $p(n)$ identical registers are needed to record all the states of the
tape during the $p(n)$ steps of DTM computation. These registers are labeled
by an integer $i$, $i=1,2,\cdots,p(n)$, corresponding to the state of the tape
right before the $i$th step of computation. The $j$th binary wire of the $i$th
register naturally gets the label $(i,j)$. Without the device called {\sc
static finite state control} (SFSC) which simulates the function of finite
state control in a static manner, the $p(n)\times p(n)$ isolated binary wires
can do nothing but record the states of the tape right before each step of DTM
computation. The SFSC is a three-in-three-out device with an EDCSQN inside,
as shown in Fig.2. The binary wires labeled by R and W are used to simulate the
read-write head. One should connect $p(n)\times p(n)$ identical SFSCs also
labeled by $(i,j)$ where $i,j=1,2,\cdots ,p(n)$ to these binary wires in the
manner that the R and W ends of the $(i,j)$th SFSC are connected to the
$(i,j)$th and $(i+1,j)$th binary wires respectively, the ID and the IU ends of
the $(i,j)$th SFSC are connected to the OD end of the $(i-1,j+1)$th SFSC and
the OU end of the $(i-1,j-1)$th SFSC respectively, with care taken to the
connection for these ends of the boundary SFSCs with $i,j=1,p(n)$. Now it is
time to explain the function of the SFSC in detail. Look at a typical
SFSC labeled by $(i,j)$. The end labeled by R ``reads'' the ``symbol'' from
the binary wire $(i,j)$ and the SFSC does some evaluation which gives a new
``symbol'' and an instruction for ``head move''. The end labeled by W
``writes'' the new ``symbol'' into the $(i+1,j)$th binary wire. And the ``head
move'' instruction is transmitted through the ends OU (outgoing-up-move) and
OD (outgoing-down-move) to the $(i+1,j+1)$th and $(i+1,j-1)$th SFSCs
respectively. Correspondingly, the IU (incoming-up-move) and ID
(incoming-down-move) ends are used to convey the ``head move'' instructions
from the former SFSCs. The function of the SFSC is controlled by the IU and
ID ends in the manner that if all the inputs from ID and IU are logic
$0$, the SFSC does nothing for the symbol on the end R, just passes it
directly to the W end, and sets both OU and OD to logic $0$. This is
simulating the tape squares which are not under the read-write head. If either
ID or IU gives logic $1$ (they never both give $1$), the SFSC does non-trivial
evaluation in simulation of the operation of the finite state control during
the $i$th step of DTM computation, outputs a new ``symbol'' through its W
end and provides ``head move'' instructions to following SFSCs. Although for
each computational stage $i$, there are as many as $p(n)$ SFSCs, only one
actually is invoked into effect of non-trivial computation, others just loyally
pass the information. Particularly, for the beginning SFSCs labeled by
$(1,j)$, $j=1,2,\cdots p(n)$, there is only one connected to the binary wire
representing the tape square at which the read-write head starts is invoked
into non-trivial effect, all the others are dormant. The SQDTM has been
constructed as some kind of static quantum automata (SQA) [9] which
``runs'' the DTM program in the static manner and outputs the results by the
last binary wires when in ground state. It is taken for granted that the
SFSC can be realized by a finite number (say $M$) of basic static logic
elements. The static computational complexity of the SQDTM, {\it i.e.} the
number of basic static logic elements it contains, is bounded by a
polynomial ${\cal P}(n)$,
\begin{eqnarray}
{\cal P}(n) &=& p(n)\times p(n)+p(n)\times p(n)\times M \nonumber \\
            &=& (M+1)p^2(n)
\end{eqnarray}
This proves
\begin{th}
{\large P $\subseteq$ SP}
\end{th}
which claims that for any problem in the class P with input size $n$, the
solution together with the inputs can be encoded into the ground state of
a static quantum computer consisting of no more than ${\cal P}(n)$ basic
static logic elements, where ${\cal P}(n)$ is a polynomial function. More
strikingly, with the help of two energy degeneracy lifting (EDL) units, a
universal static quantum computer (USQC) can be constructed to encode the
solutions of any NP problem into the ground state.
\begin{df}
A {\sc decision energy degeneracy lifting unit} (DEDLU) is simply a
binary wire with one end inside the unit subject to a bias field that
lifts the energy degeneracy between the two working modes of the binary
wire and the other end called {\sc decision port} (DPORT) connected
externally.
\end{df}
\begin{df}
A {\sc minimization energy degeneracy lifting unit} (MEDLU) is actually
a group of binary wires each of which has one end inside the unit
subject to an appropriate bias field. The other ends form the group called
{\sc minimization port}s (MPORTs) are for external connection. The bias
fields are adjusted so that when the minimization ports are assigned binary
values $z_0,z_1,\cdots ,z_m$ respectively that may be used to denote an
integer
\begin{equation}
Z=\sum _{i=0}^{m}z_i2^i
\end{equation}
the total interaction energy inside the MEDLU should be directly
proportional to the number $Z$.
\end{df}
\begin{df}
A {\sc universal static quantum computer} can be constructed from an EDC
SQDTM with some DEDLUs and a MEDLU connected to the output, as shown in
Fig.3.
\end{df}

By definition, an NP decision problem is a problem whose answer
can be checked in polynomial time on a classical DTM [5]. In general, the
check consists of a few YES/NO questions. According to Theorem 1, this
check task can be accomplished by an EDC SQDTM
consisting of polynomial number of basic static logic elements which is part
of the USQC as shown in Fig.3. By virtue of the static quantum parallelism,
all instances of this search problem are stored in the input binary wires,
the EDC SQDTM checks each instance and gives YES (say logic $1$) or NO (say
logic $0$) answers to the output binary wires which connect with the DPORTs
of the DEDLUs. Without the EDL units, all the instances are still degenerate.
While a DEDLU lifts the energy degeneracy by providing a higher or lower
interaction energy according to the logic value on the DPORT is $0$ or $1$.
Therefore instances which lead to all YES answers will keep the USQC in the
ground state while those which lead to at least one NO answer will lift
the USQC to higher energies. By this means the USQC does the nondeterministic
computations and labels the YES and NO instances by lower and higher total
energies. When the computer is surely in the ground state, a measurement gives
the answer of the problem. The USQC further contains a MEDLU intended for
minimization search problems. The quantity to be minimized is calculated for
each instance and the results are sent to the MPORT. The MEDLU in turn gives
an interaction energy proportional to these values as  described in
Definition 10. The ground state of the USQC encodes the instance which
minimize the objective quantity. Actually the USQC can be used to solve all
optimization problems since any of them can be easily transformed into a
minimization problem. According to the definitions, the two EDL units simply
consist of binary wires subject to appropriate bias fields. Since the number
of output wires of the EDC SQDTM is polynomially bounded, it is obvious that
all of the EDLs can be constructed with the total static complexity bounded by
a polynomial function of the problem input size. What has been proved is the
fact that any NP problem can be ``solved'' on a universal static quantum
computer consisting of polynomial number of basic static logic elements, in
the sense that the solutions are encoded in the ground state of the computer,
in mathematical form
\begin{th}
{\large NP $\subseteq$ SP}.
\end{th}

\begin{center}
{\bf Conclusions}
\end{center}

On a universal static quantum computer, all P and NP problems can be
efficiently encoded in the sense that the number of logic gates, the
precision in energy tailoring are all bounded by polynomial functions of
the problem input size. Unfortunately, SP does not imply a polynomial
solution of an NP problem. We still don't know how to make the system go
fast to its ground state. The usual way of relaxation often gets slow
exponentially. Nevertheless, one may use a static quantum computer to
simulate the dynamcis of other complex systems and it should be interesting
to study how the ``hardness'' of an NP problem is related to the slow
relaxation of a correspondingly designed physical system.
On the technological side, nanotechnology holds the promise to realize the
desired strongly interacting quantum systems. It seems possible to realize
SQLGs and SQNs using superconductive devices [10,11] with today's
well-established technology.

\begin{center}
{\large FIGURE CAPTIONS}
\end{center}

\noindent
Fig.1\hspace{5 mm}The Symmetrized AND Gate  

\noindent
Fig.2\hspace{5 mm}The Static Finite State Control

\noindent
Fig.3\hspace{5 mm}The Universal Static Quantum Computer 

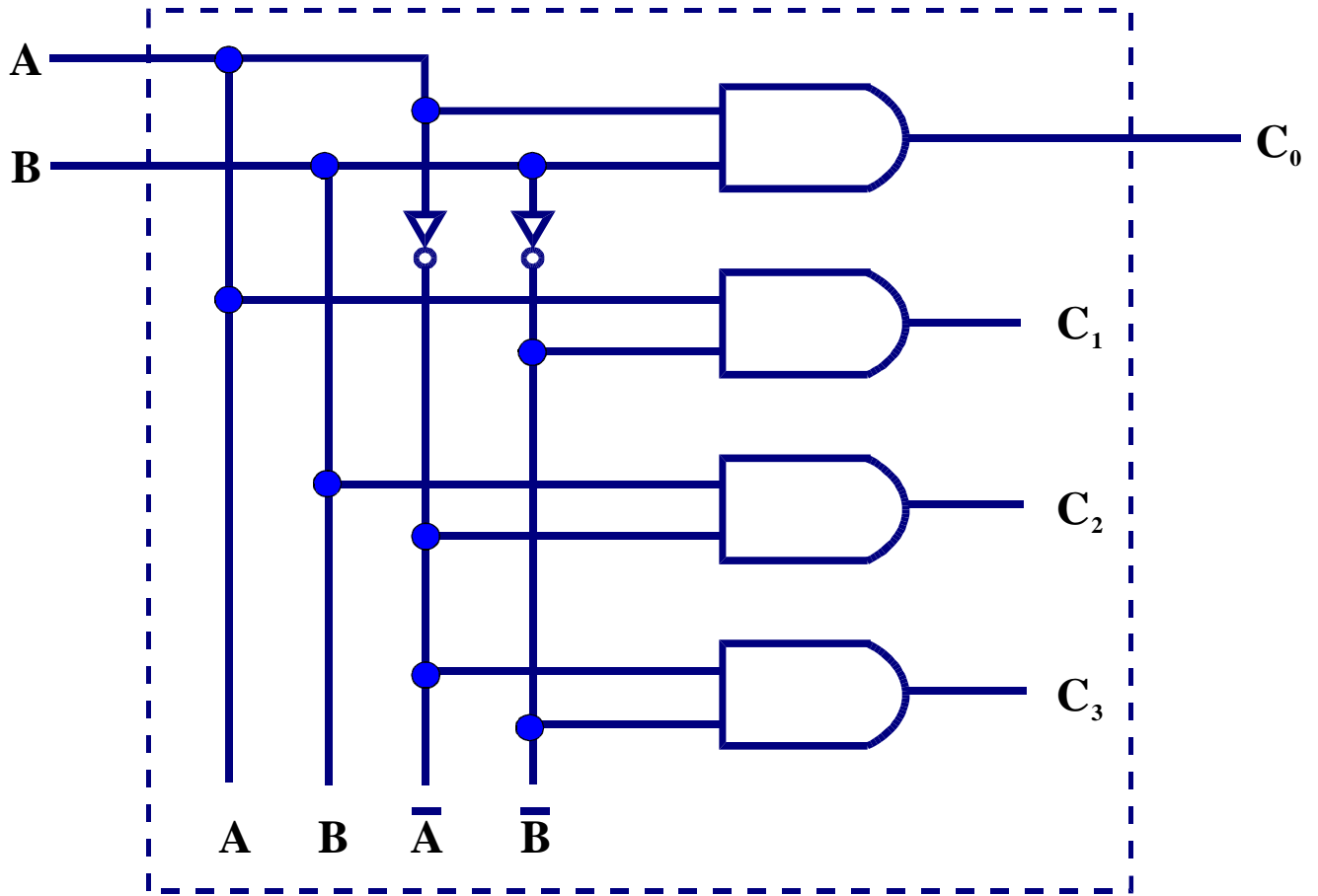
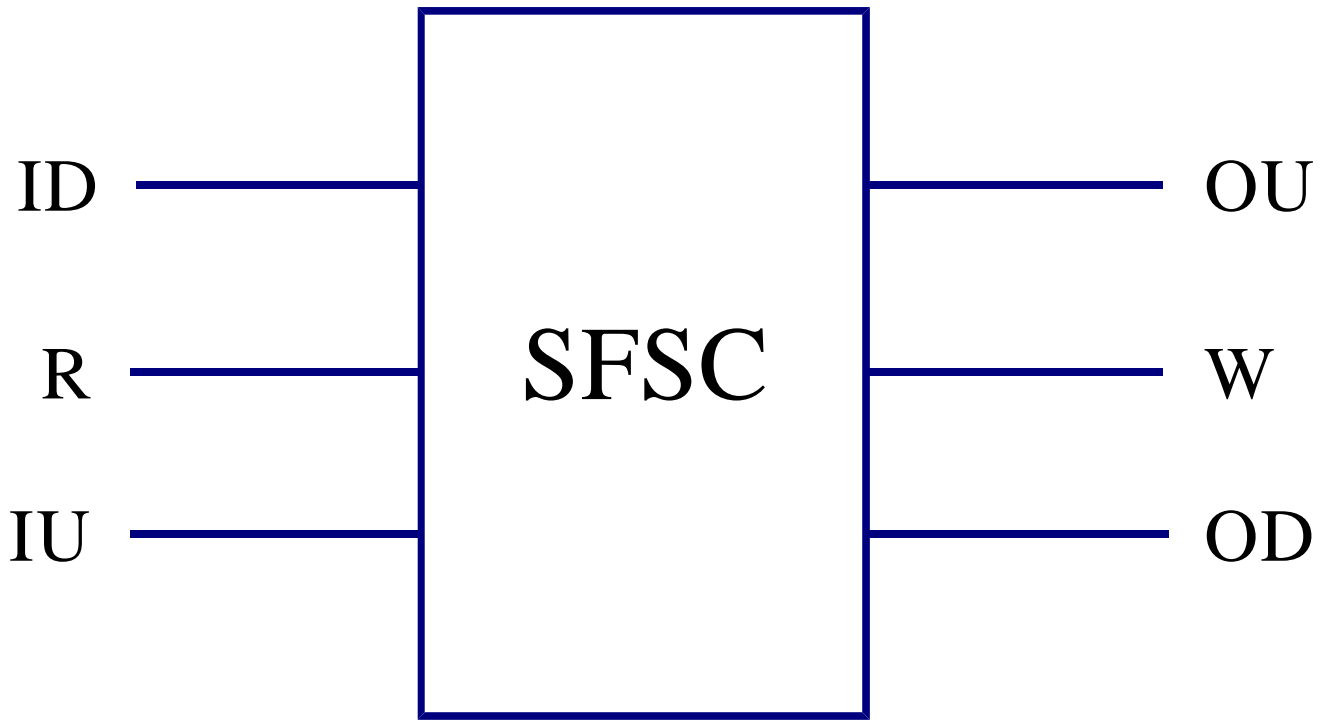
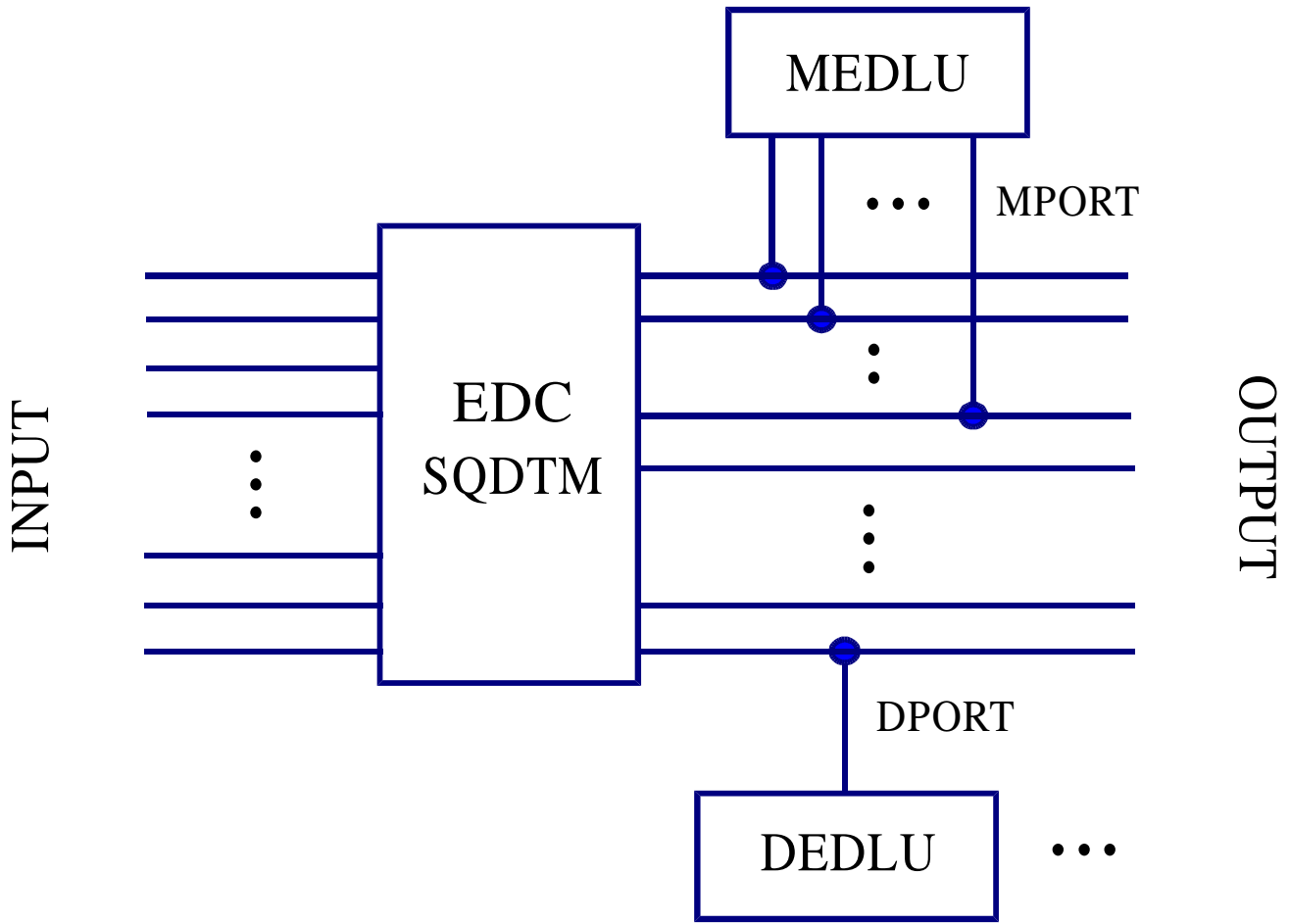
\end{document}